\documentclass{ws-ijmpa}

\begin{document}

\markboth{B.A. Arbuzov and I.V. Zaitsev}
{Upgraded LHC experiments and SM fundamental
parameters }

%
%

\title{On a possible effective four-boson interaction and its
implications at the upgraded LHC}

\author{Boris A. Arbuzov}

\address{\it M.V. Lomomnosov Moscow State University,\\
119991 Moscow, Russia\\
arbuzov@theory.sinp.msu.ru}

\author{Ivan V. Zaitsev}

\address{\it M.V. Lomomnosov Moscow State University,
\\119991 Moscow, Russia\\
zaitsev@theory.sinp.msu.ru}

\maketitle

\begin{history}
\received{Day Month Year}
\revised{Day Month Year}
\end{history}

\begin{abstract}
We consider a possibility of a spontaneous
generation of fourfold effective
interactions of electroweak gauge bosons $W^a$ and $B$.  The conditions for the
spontaneous generation are shown to lead to a
set of compensation equations for parameters of the interaction.
In case of a realization of a non-trivial solution of the set, important electro-weak parameter $\sin^2\theta_W$ is defined. The existence
of two non-trivial solutions
is demonstrated, which provide a satisfactory value for
the electromagnetic fine structure constant $\alpha$ at scale
$M_Z$:
$\alpha(M_Z) = 0.007756$.
There is a solution with the high effective cut-off being close to
the Planck mass by the order of magnitude.
The most interesting solution corresponds to effective cut-off
$\simeq 10^2\,TeV$. This solution gives quite definite prediction for
non-perturbative effects
in processes $p+p\to\bar t t (W^\pm, Z)$, which could be observed
at the upgraded LHC.
 \end{abstract}

\ccode{PACS numbers: 12.15.-y, 12.15.Ji, 14.70.Fm, 14.80.Ec}
\bigskip
\section{Introduction}
The Standard Model of particles' interactions is fairly considered
as quite successful theory. It explains the phenomena in high
energy physics experiments, and give us consecutive interpretation
of their
totality structure. But till now we cannot tell that SM is
all-sufficient theory. First of all, by the reason of that it
cannot explain at least lowest energy gravitational interaction
in coupling with other ones and in the same way as others. Just it
cause theorists to make attempts of different SM extensions
building.

However, secondly, even if we shall leave such problems as scale
hierarchy aside, we immediately face another, so to say, more
prosaic problem. We cannot admit the Standard Model as
accomplished theory simply by the reason of that there are too many
external parameters we have to bring into it for provide it's
expository
power. The number of these parameters (such as coupling constants
and matter fields' masses ratios) reaches up to 29. Of course, we can
hope that determination of their value will be supplied with the
mentioned wouldbe extension of the SM.
But we see striking contrast between existing theory's insularity
and it's wide experimental validation just in this insularity,
on the one hand, - and  further developed constructions' not fixed
status with the absence of any data confirmation at present day and indeterminate perspectives in the foreseeable future, on the  other
hand.

Provided the wouldbe extended theory will be really able to reduce
the totality of all data
to one general principle, we have to make extremely great progress
in experiment technology for making this theory as such
validated as SM is validated just now.
This problem prompts us to make efforts in other direction.
Particularly, we can attempt to lead out necessary  evaluations
just within the existing theory structure. And those minimal
extensions which we'll have to build anyway, must rather be
non-structural and deal not with new fields and particles, but with
new type of just known ones' effective interactions.
And the searching region for such interactions may be, of course,
indicated by the fact that SM and general quantum field theory
conception are of the less successfulness in describing of lower
energy processes. We can ask ourselves: may be, there is some deep
correlation
between the fact of failing of perturbation theory in such
phenomena description and the presence of special low energy
quantum effects, which are to be taken into account in some way?
This situation patently corresponds
with such effects as superconductivity and superfluidity, where
classical local theory was powerless and consecutive analysis in
terms of fundamental quantum theory equations was inaccessible
also, but where the solution was found in the framework of
non-perturbative contributions. We have no perturbative source of
a "force", which bind electrons into
Cooper's pair, but we can describe their behavior in this
pairing.

The method of effective non-local interactions building,
which we shall try to apply to the mentioned above problem in this
work, was grown up from N.N. Bogolyubov's
compensation conception \cite{1,2} developed
and successfully applied just in the superconductivity theory.
Although in field theory it acquire some new specialities,
stated above analogy seems to be quite encouraging for us.
And on the other hand, we've just also quite successfully
applied this approach to the range of particle physics low energy
processes. The compensation approach was applied \cite{3,4}
to the problem of the spontaneous generation of effective
interactions in quantum field theories. Most impressive
effectiveness of the method was demonstrated
in light meson physics, where spontaneously generated
Nambu - Jona-Lasinio interaction \cite{5,6} building
allowed us \cite{7} to predict
main particles' properties with good precision using only
fundamental QCD parameters without  external
parameters bringing in. Also applications to the composite Higgs
particle problem \cite{8} and to the spontaneous generation of
the wouldbe anomalous three-boson interaction \cite{9,10},
to be discussed below, can be mentioned.
Our aim in this work is to demonstrate principal possibility of
finding solution for fundamental SM parameters problem
in terms of effective interactions. Correspondingly, we build
a simple model, being guided by our previous experience in
similar, but more advanced models. In case of success
of this attempt it would be really important step, hopefully opening a
road to the more sophisticated and more close to reality theorizing
upon this important subject. But, in the same time, just with
this
simplified approach we shall present some predictions, suitable for
upgraded LHC experiments studies.

In works \cite{4,5,6,7,8,9,10,11}
N.N. Bogoliubov compensation principle \cite{1,2}
was applied to studies of a spontaneous generation of effective
non-local interactions in renormalizable gauge theories. The method
and applications are also
described in full in the book \cite{12}.

In particular, papers \cite{9,10}  deal with an application of the approach to the electro-weak interaction and a possibility of spontaneous generation of effective anomalous three-boson interaction of the form
\begin{eqnarray}
& &-\,\frac{G}{3!}\,F\,\epsilon_{abc}\,W_{\mu\nu}^a\,W_{\nu\rho}^b\,
W_{\rho\mu}^c\,;
\label{FFF}\\
& &W_{\mu\nu}^a\,=\,
\partial_\mu W_\nu^a - \partial_\nu W_\mu^a\,+g\,\epsilon_{abc}
W_\mu^b W_\nu^c\,.\nonumber
\end{eqnarray}
with uniquely defined form-factor $F(p_i)$, which guarantees effective interaction~(\ref{FFF}) acting in a limited region of the momentum space. It was done in the framework of an approximate scheme, which accuracy was estimated to be $\simeq (10 - 15)\%$ \cite{3}. Would-be existence of effective interaction~(\ref{FFF}) leads to important non-perturbative
effects in the electro-weak interaction. It is
usually called anomalous three-boson interaction and it is considered
for long time on phenomenological grounds \cite{13,14}. Our
interaction constant $G$ is connected with
conventional definitions in the following way
\begin{equation}
G\,=\,-\,\frac{g\,\lambda}{M_W^2}\,;\label{Glam}
\end{equation}
where $g \simeq 0.65$ is the electro-weak coupling.
The best limitations for parameter $\lambda$ read \cite{15}
\begin{equation}
\lambda_\gamma = -\,0.022\pm0.019\,;\quad
 \lambda_Z = -\,0.09\pm0.06\,; \label{lambda1}
\end{equation}
where subscript denote a neutral boson being involved in the
experimental definition of $\lambda$.

For the electro-weak interaction we have \cite{9,10} as conditions for a
spontaneous generation of interaction (\ref{FFF}) following relations
\begin{equation}
g(z_0) = 0.60366\,;\; z_0 = 9.6175\,;\;  G\,=\,0.000352\,TeV^{-2}.\label{eq:gz0}
\end{equation}
Here $z_0$ is a dimensionless parameter, which is connected with
value of a boundary momentum, that is with effective cut-off
$\Lambda_0$ according  to the following definition~\cite{9, 10}
\begin{equation}
\frac{2\,G^2\,\Lambda_0^4}{1024\,\pi^2}\,=\,
\frac{g^2\,\lambda^2\,\Lambda_0^4}{512\,\pi^2\,M_W^4}\,=\, z_0 \,.\label{eq:Lambda}
\end{equation}
Let us note, that the solution of the analogous compensation
procedure in QCD correspond to  $g(z_0)=3.817$~\cite{11}, that
gives satisfactory description of the low-momentum behavior of
the running strong coupling.

\begin{figure}
\includegraphics[scale=0.55]{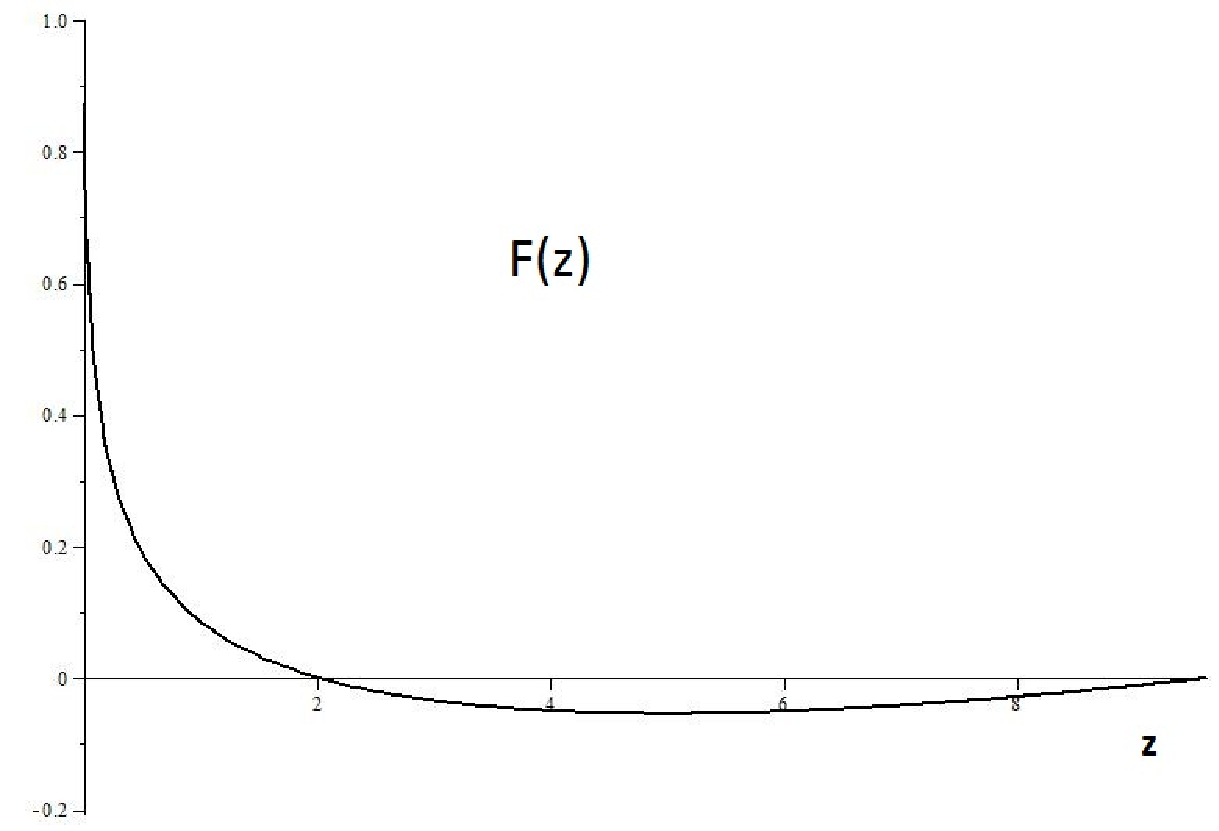}
\caption{The behavior of the form-factor for the electro-weak theory.}
\label{fig:compenG}
\end{figure}

It is instructive to present in Fig.~\ref{fig:compenG} the behavior of form-factor $F(p,-p,0)$ in dependence on momentum $p$, where
\begin{equation}
z\,=\,\frac{G^2\,p^4}{512\,\pi^2}\,;\label{eq:zdef}
\end{equation}
and $F(z)\,=\,0$ for $z\,>\,z_0$.
As a rule the existence of a non-trivial solution of a compensation
equation impose essential restrictions on parameters of a problem. Just
the example of these restrictions is the definition of coupling constant
$g(z_0)$ in~(\ref{eq:gz0}). It is advisable to consider other
possibilities
for spontaneous generation of effective interactions and to find out,
which restrictions on physical parameters may be imposed by an existence
of non-trivial solutions. In the present work we consider
possibilities of definition of links between important physical
parameters, first of all with relation to the fine structure
constant $\alpha$.

\section{Weinberg mixing angle and the fine structure constant}
Let us demonstrate a simple model, which illustrates how the well-known Weinberg mixing angle
could be defined. Let us  consider a possibility of a spontaneous
generation of the following
effective interaction of electroweak gauge bosons
\begin{eqnarray}
& & L_{eff}^{W} =
-\,\frac{G_2}{8} W^a_\mu W^a_\mu W^b_{\rho \sigma}
W^b_{\rho \sigma} - \frac{G_3}{8}\,W^a_\mu W^a_\mu B_{\rho \sigma}
B_{\rho \sigma} -\nonumber\\
& &\frac{G_4}{8}\,Z_\mu Z_\mu W^b_{\rho \sigma} W^b_{\rho \sigma}
- \frac{G_5}{8}\,Z_\mu Z_\mu B_{\rho \sigma} B_{\rho \sigma} .
\label{eq:LeffWZ}
\end{eqnarray}
where we maintain the residual gauge invariance for the electromagnetic
field. Here indices $a,d$ correspond to charged $W$-s, that is they
take
values $1,\,2$, while index $b$
corresponds to three components of $W$ defined by the initial
formulation
of the electro-weak interaction. Definition~(\ref{eq:LeffWZ})
corresponds to convenient rule for Feynman rules
for corresponding vertices, {\it e.g.} for the first term in~
(\ref{eq:LeffWZ})
\begin{equation}
\imath\,G_2\,g_{\mu \nu}\,(g_{\rho \sigma} (p\,q)\,-\,p_\sigma\,
q_\rho)\;
\label{FRV}
\end{equation}
where components of $W^a$ have indices $\mu,\,\nu$ and incoming momenta
and indices $(p,\rho)$ and $(q, \sigma)$ refer to fields
$W^b$.
Let us remind the relation, which connect fields $W^0,\,B$ with physical
fields of the $Z$ boson
and of the photon
\begin{eqnarray}
& &W^0_\mu = \cos \theta_W\,Z_\mu + \sin \theta_W\,A_\mu ;\nonumber\\
& &B_\mu = -\,\sin \theta_W\,Z_\mu + \cos \theta_W\,A_\mu .
\label{eq:Wmix8}
\end{eqnarray}
Thus in terms of the physical states ($W^+\, W^-\, Z\, A\,$)
wouldbe effective interaction~(\ref{eq:LeffWZ}) has the following
form
\begin{eqnarray}
& &L_{eff}^W\,=\,-\frac{G_2}{2}\,W_\mu^+\,W_\mu^-\,W^+_{\rho\sigma}
\,W^-_{\rho\sigma}\,-
\frac{G_2}{4}\,W_\mu^+\,W_\mu^-\,\Bigl(\cos^2\theta_W\,
Z_{\rho\sigma}\,Z_{\rho\sigma}\,+\nonumber\\
& &2\,\cos\theta_W\sin\theta_W\,
Z_{\rho\sigma}\,A_{\rho\sigma}\,+\,\sin^2\theta_W\,A_{\rho\sigma}
\,A_{\rho\sigma}\Bigr)\,-\frac{G_4}{4}\,Z_\mu\,Z_\mu\,W^+_{\rho\sigma}
\,W^-_{\rho\sigma}\,-\label{eq:intG}\\
& &\frac{G_4}{8}\, Z_\mu Z_\mu \Bigl(\cos^2\theta_W
Z_{\rho\sigma} Z_{\rho\sigma} + \sin^2\theta_W A_{\rho\sigma}
A_{\rho\sigma} +2 \cos\theta_W\sin\theta_W
Z_{\rho\sigma} A_{\rho\sigma}\Bigr)-\nonumber\\
& &\frac{G_3}{4}\, W^+_\mu W^-_\mu \Bigl(\sin^2\theta_W
Z_{\rho\sigma} Z_{\rho\sigma} + \cos^2\theta_W A_{\rho\sigma}
A_{\rho\sigma} -2 \cos\theta_W\sin\theta_W
Z_{\rho\sigma} A_{\rho\sigma}\Bigr)-\nonumber\\
& &\frac{G_5}{8}\, Z_\mu Z_\mu \Bigl(\sin^2\theta_W
Z_{\rho\sigma} Z_{\rho\sigma} + \cos^2\theta_W A_{\rho\sigma}
\,A_{\rho\sigma}\,-2\,\cos\theta_W\sin\theta_W\,
Z_{\rho\sigma} A_{\rho\sigma}\Bigr).\nonumber
\end{eqnarray}

Interactions of type~(\ref{eq:intG}) were earlier introduced on phenomenological
grounds in works \cite{16,17} and are subjects for experimental
studies.
Let us introduce an effective cut-off $\Lambda$
and consider a possibility of a spontaneous generation of interaction~(\ref{eq:LeffWZ}).
In doing this we proceed with the
add-subtract procedure, which was used throughout works~
\cite{3,4,5,6,7,8,9,10}. Now we start with
usual form of the Lagrangian, which describes electro-weak gauge
fields $W^a$ and $B$
\begin{eqnarray}
& &L = L_0\,+\,L_{int}\,;\nonumber\\
& &L_0 = -\,\frac{1}{4}\bigl(
W_{0\mu\nu}^a\,W_{0\mu\nu}^a\bigr)\,-\,\frac{1}{4}\bigl( B_{\mu\nu}\,B_{\mu\nu}\bigr)\,;\label{eq:L0WB}\\
& &L_{int}\,=\,-\frac{1}{4}\bigl( W_{\mu\nu}^a\,W_{\mu\nu}^a-W_{0\mu\nu}^a\,W_{0\mu\nu}^a\bigr) \,.\label{eq:LIntWB}\\
& &W_{0\mu\nu}^a = \partial_\mu W_\nu^a - \partial_\nu W_\mu^a;\;
B_{\mu\nu} = \partial_\mu B_\nu - \partial_\nu B_\mu .\nonumber
\end{eqnarray}
and $W_{\mu\nu}^a$ is the well-known non-linear Yang-Mills field of
$W$-bosons.
Then we perform the add-subtract procedure of
expression~(\ref{eq:LeffWZ})
\begin{eqnarray}
& &L = L'_0\,+\,L'_{int}\,;\nonumber\\
& &L'_0 = L_0\,-\,L_{eff}^{W}\,; \label{eq:L'WB0}\\
& &L'_{int}\,=\,L_{int}\,+\,L_{eff}^{W}\,. \label{eq:L'WBInt}
\end{eqnarray}

\begin{figure}
\includegraphics[scale=0.45]{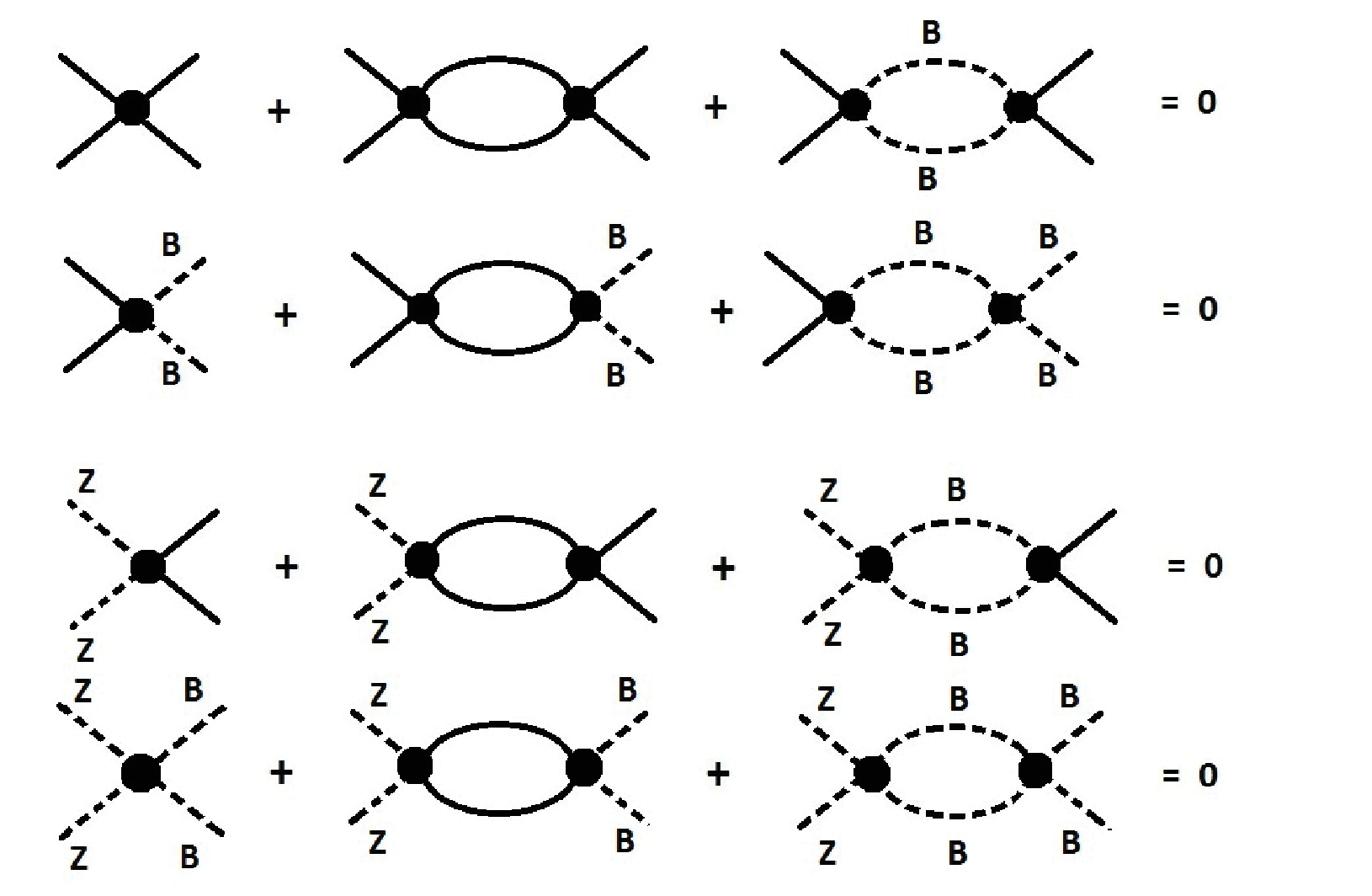}
\caption{Diagram representation of set~(\ref{eq:CompWB}).
Simple line represent
$W$-s, dotted lines represent $B$ and lines, consisting of black spots, represent $Z$.}
\label{fig:NJLBS}
\end{figure}

Now let us formulate compensation equations for wouldbe interaction~(\ref{eq:LeffWZ}).
We are to demand, that considering the theory with Lagrangian
$L'_0$~(\ref{eq:L'WB0}), all contributions to four-boson
connected vertices, corresponding to interaction~(\ref{eq:LeffWZ}) are
summed up to zero. That is the undesirable interaction part in the
would-be free Lagrangian~(\ref{eq:L'WB0}) is compensated. Then we are
rested with
interaction~(\ref{eq:LeffWZ}) only in the proper
place~(\ref{eq:L'WBInt})
We would formulate these compensation equations using experience
acquired in the course of application of the method to the Nambu -
Jona-Lasinio interaction and the triple weak boson
interaction~(\ref{FFF}). As is demonstrated in book~\cite{12}
(Section 3.3), the first approximation for the problem of
spontaneous generation of the
Nambu - Jona-Lasinio interaction takes form-factor $F(p)$ to be
a step function $\Theta(\Lambda^2 -p^2)$ and
only horizontal diagrams of the type presented in Fig.~\ref{fig:NJLBS}
are taken
into account. The next approximation, described in detail
in \cite{4} and in \cite{12} (Chapter 5) includes also
vertical diagrams and form-factor $F(p)$ is uniquely defined as a
solution of the set of the compensation conditions in terms of standard
Meijer functions. We have
demonstrated, that the first approximation gives satisfactory results
and the next ones serves for its specification. In the
present work we just use the first approximation.

So let us introduce effective cut-off $\Lambda$, which is a
subject for definition by solutions of the problem and use
just a step function $\Theta(\Lambda^2 -p^2)$ for the effective
form-factor.

In this way we have the following set of compensation equations, which corresponds to diagrams being presented in
Fig.~\ref{fig:NJLBS}
\begin{eqnarray}
& &-\,x_2\, -\,2\,F_W\,x_2^2\,-\,(1-a^2)\, F_Z\, x_3\, x_4\,-
a^2\,F_Z\,x_2\, x_4\,=\,0\,;\label{eq:CompWB}\\
& &-\,x_3\,-\,2\,F_W\,x_2\,x_3\,-\,a^2\,F_Z\,x_2 x_5-(1-a^2)F_Z\,
x_3\,x_5\,=0\,;\nonumber\\
& & -\,x_4\,-\,2\,F_W\,x_2\, x_4\,-\,a^2\,F_Z\,x_4^2\,-\,(1-a^2)\,
F_Z x_3\,x_4\,=\,0\,;\nonumber\\
& &-\,x_5\,-\,2 \,F_W\,x_3\,x_4\,-a^2\,F_Z\,x_4\,x_5\,
-\,(1-a^2)\,F_Z\,x_5^2
\,=\,0\,;
\nonumber\\
& &F_W\,=\,1\,-\,\frac{2 M_W^2}{\Lambda^2}\Bigl(L_W-\frac{1}{2}\Bigr);\nonumber\\
& &F_Z\,=\,1\,-\,\frac{2 M_Z^2}{\Lambda^2}\Bigl(L_Z-\frac{1}{2}\Bigr)\,;\nonumber\\
& &x_i\,=\,\frac{3\,G_i\,\Lambda^2}{16\,\pi^2}\,;\;L_W\,=\,\ln \frac{\Lambda^2+M_W^2}{M_W^2}\,;\;L_Z\,=\,\ln
\frac{\Lambda^2+M_Z^2}{M_Z^2}\,;\;
a\,=\,\cos\theta_W.\nonumber
\end{eqnarray}
Factor $2$ in equations here corresponds to sum by weak
isotopic index $\delta^a_a\,=\,2,\,a = 1,\,2$.

We have the following solutions of set~(\ref{eq:CompWB}) in addition
to the evident trivial one: $x_2=x_3=x_4=x_5=0$
\begin{eqnarray}
& &x_3\,=\,x_5\,=\,0;\;x_2\,=\,-\,\frac{1+a^2 F_Z x_4}{2\,F_W};
\label{sol1}\\
& &x_3\,=\,x_5\,=\,0;\;x_2\,=\,-\,\frac{1}{2\,F_W};\;x_4\,=\,0;
\label{sol2}\\
& &x_2\,=\,x_4\,=\,0;\;x_3\,=\,\frac{a^2}{2(1-a^2)F_W};\;
x_5\,=\,-\,\frac{1}{(1-a^2)F_Z};
\label{sol3}\\
& &x_2\,=\,x_4\,=\,-\,\frac{1}{2\,F_W};\;x_3\,
=\,\frac{a^2}{2(1-a^2)F_W};\,
x_5\,=\,-\,\frac{1}{(1-a^2)F_Z};
\label{sol4}\\
& &x_2\,=\,-\,\frac{1}{2\,F_W};\;\,x_4\,=\,0;\,
x_3\,=\,0;\,
x_5\,=\,0;
\label{sol5}\\
& &x_2\,=\,x_4\,=\,0;\;
x_5\,=\,-\,\frac{1}{(1-a^2)F_Z};
\label{sol6}\\
& &x_2\,=\,-\,\frac{1}{2\,F_W};\;\,x_4\,=\,0;\,
x_3\,=\,\frac{a^2}{2(1-a^2)F_W};\,
x_5\,=\,-\,\frac{1}{(1-a^2)F_Z};
\label{sol7}\\
& &x_2\,=\,-\,\frac{1}{2\,F_W};\;\;x_4\,=\,0;\;x_5\,=\,0;
\label{sol8}\\
& &x_2\,=\,x_4\,=\,-\,\frac{1+(1-a^2)F_Z\,x_5}{2\,F_W+a^2\,F_Z};\,
x_3\,=\,x_5;\label{sol9}
\end{eqnarray}
Note, that absence of some $x_i$ in a solution means that this $x_i$ is arbitrary.

\begin{figure}
\includegraphics[scale=0.32]{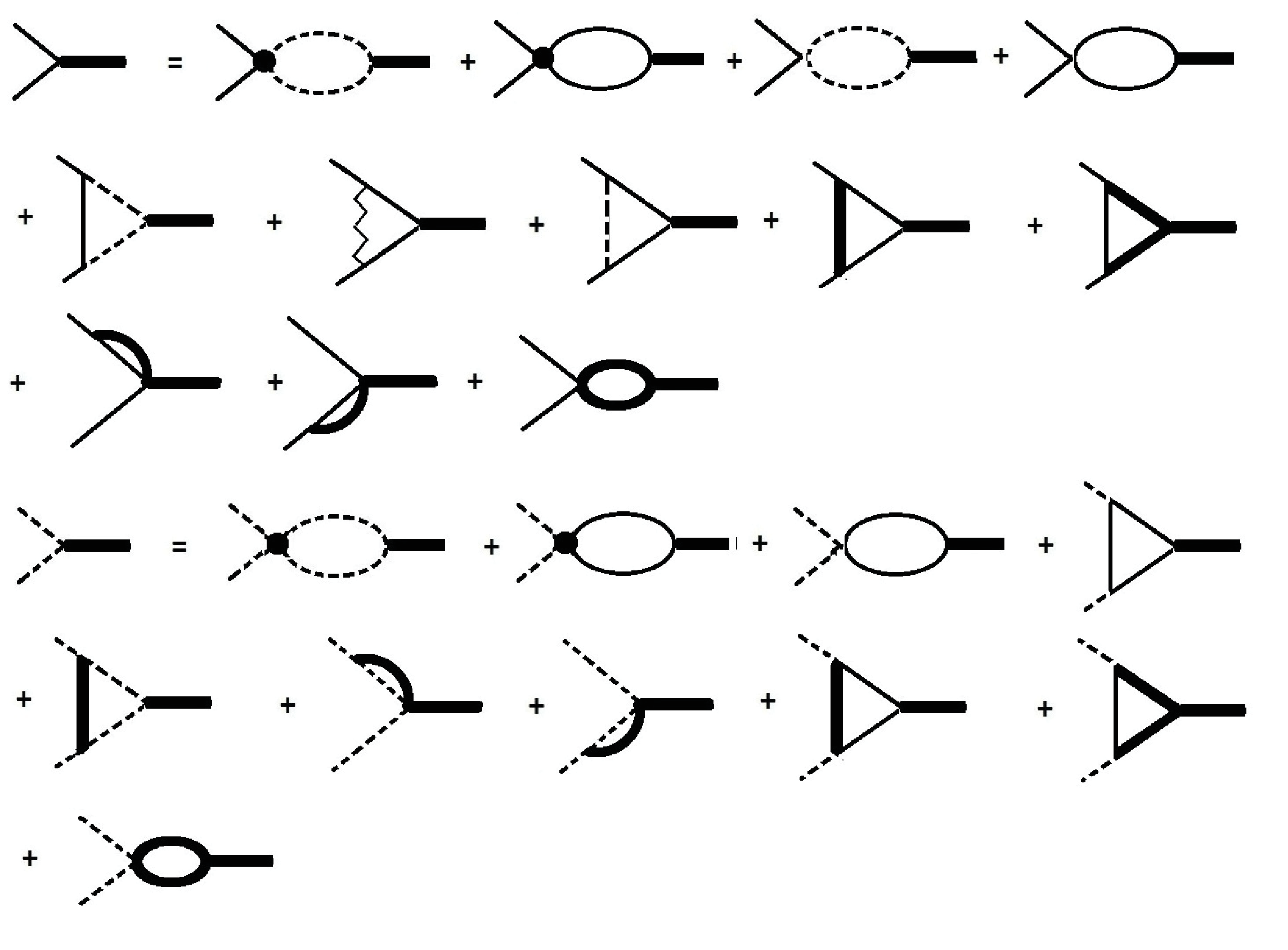}
\caption{Diagram representation of set~
(\ref{BSW2},\ref{BSW4}).
Simple line represent
$W$-s, dotted lines represent $B$, wave line represents a photon and lines, consisting of black spots, represent $Z$.  Thick lines represent the Higgs scalar.}
\label{fig:NJLBS3}
\end{figure}
Then, following the reasoning of the approach, we assume, that
the Higgs scalar corresponds to a bound state consisting  of a
complete set of fundamental particles. Here we study the wouldbe
effective interaction~(\ref{eq:LeffWZ}, \ref{eq:intG}) of the
electroweak
bosons, so we take into account just these bosons as constituents
of the Higgs scalar.
There are two Bethe-Salpeter equations for this bound state,
because constituents are either $W^a\,W^a$ or $Z\,Z$.
These equations are presented in the two rows of
Fig.~\ref{fig:NJLBS3}. In approximation of very large cut-off
$\Lambda$ these equations have the following form with notations
of~(\ref{eq:CompWB})
\begin{eqnarray}
& &-\,3\, x_2\, (2\, F_W+a\, F_Z) -\,\frac{x_3(1-a^2)}{a} -
\frac{3\, \alpha_{ew}}{16\, \pi}
\Bigl[- \frac{a^2(a^6-a^4-5\, a^2+1)}{1-a^2}L_W+ \nonumber\\
& &\frac{(1+a^2)(1-3\, a^2)}{a^2(1-a^2)}L_Z-
\frac{(1-a^2-a^4)(1-a^2)}{a^2}\Bigr]+\label{BSW2}\\
& &\frac{3 \,\alpha_{ew}\, M_W^2}{32\, \pi}\Bigl[\frac{3\, M_H^2 }{(M_H^2-M_W^2)^2}\ln\Bigl[\frac{M_H^2}{M_W^2}\Bigr]-
\frac{3}{M_H^2-M_W^2}-\frac{8}{M_W^2}\Bigl] =
\frac{1}{B_W};\nonumber\\
& &-\,x_4\,(2\,F_W +a\,F_Z) -\frac{x_5\,(1-a^2)}{a}
- \frac{\alpha_{ew}\,a^2}{4 \pi}+\nonumber\\
& &\frac{3 \,\alpha_{ew}\, M_Z^2}
{32\, \pi a^4}\Bigl[\frac{3\, M_H^2 }{(M_H^2-M_W^2)^2}\ln\Bigl[\frac{M_H^2}{M_Z^2}\Bigr]-\frac{3}
{M_H^2-M_Z^2}-\frac{8}{M_Z^2}\Bigl]
  =
\frac{1}{a^2 B_Z};\label{BSW4}\\
& &
B_W=F_W+\frac{ M_H^2}{2\,\Lambda^2}\Bigl(L_W-\frac{13}{12}\Bigr);
\;B_Z=F_Z+\frac{ M_H^2}{2\,\Lambda^2}\Bigl(L_Z-\frac{13}{12}\Bigr);
\nonumber\\
& &\alpha_{ew} = \frac{\alpha_0}{1+\frac{5\alpha_0}{4 \pi} \ln\frac{\Lambda^2}{M_Z^2}};\; \alpha_0=0.0337;\;
a = \cos\theta_W(\Lambda);
\label{eq:a}\\
& &1-a^2=\frac{\alpha\Bigl(1+\frac{5 \alpha_0}{6 \pi}\ln\frac{\Lambda^2}{M_Z^2}\Bigr)}{\alpha_0\Bigl(1-\frac{5 \alpha}
{6\pi}\ln\frac{\Lambda^2}{M_Z^2}\Bigr)}\;;\;
\alpha=\frac{e^2(M_Z)}{4\,\pi} = \alpha(M_Z)=0.007756\,.
\nonumber
\end{eqnarray}

Now we look for solutions of set~(\ref{eq:CompWB},
\ref{BSW2}, \ref{BSW4}, \ref{eq:a}) for
variables $x_2,\,x_3,\,x_4,\,x_5,\, \Lambda$, which gives
appropriate value for $\alpha(M_Z)=0.007756$, according to relation~(\ref{eq:a}).
We use values for physical masses
\begin{equation}
M_W=0.0804\,TeV,\quad M_Z=0.0912\,TeV,\quad M_H=0.1251\,TeV.\label{mass}
\end{equation}
We have studied solutions of set of equation and have come to
the conclusion, that only solutions~(\ref{sol1}), (\ref{sol6}) and~(\ref{sol9})
of compensation
set~(\ref{eq:CompWB}) gives necessary value $\alpha(M_Z)=0.007756$.
For the first option~(\ref{sol1}) there are two solutions which
satisfy our conditions. Namely, the
following ones, where $a$ and $x_4$ are just solutions of the
set and $x_2$ is defined by relation~(\ref{sol1})
\begin{eqnarray}
& &\Lambda = 5.226\cdot 10^{2}\,TeV;\;x_2 = -\,0.3238;\;x_4 =
-\,0.4865\,; a = 0.8511\,;\label{solutionlow}\\
& &\Lambda = 8.687\cdot10^{16} TeV;\,x_2 = -\,0.3160;\,
x_4=-\,0.7113;\; a = 0.7192\,; \label{solutionhigh}
\end{eqnarray}
These solutions define coupling constants of effective
 interaction~(\ref{eq:LeffWZ}) again for the two solutions
\begin{eqnarray}
& &G_2 = -6.24\cdot 10^{-5}\,TeV^{-2};\;G_4 = -9.376\cdot 10^{-5}\,TeV^{-2}\,;\label{solutionlowG}\\
& &G_2 = -2.2045\cdot 10^{-33}\,TeV^{-2};\;G_4 = -4.962\cdot 10^{-33}\,TeV^{-2}\,. \label{solutionhighG}
\end{eqnarray}
From definition of parameters in experimental work \cite{19}
\begin{equation}
L_{eff}\,=\,-\,\frac{e^2 a_0^W}{8 \Lambda'^2}A_{\mu \nu}
A_{\mu \nu}W^+_\rho
W^-_\rho - \,\frac{e^2 g^2 k_0^W}{\Lambda'^2}\,A_{\mu \nu}
Z_{\mu \nu}W^+_\rho W^-_\rho\,;
\label{Leff}
\end{equation}
and from~(\ref{eq:LeffWZ}) we have
\begin{equation}
\frac{a_0^W}{\Lambda'^2} = \frac{2\,G_2}{g^2};\quad \frac{k_0^W}{\Lambda'^2}=\frac{G_2 \cos\theta_W}
{2\, g^4 \sin\theta_W}.\label{ak}
\end{equation}
Results~(\ref{solutionlowG}, \ref{solutionhighG}) lead to
the following prediction
for parameters $a_0^W,\,k_0^W$ for the two solutions
\begin{eqnarray}
& &\frac{a_0^W}{\Lambda'^2} = -0.000147\,TeV^{-2}\,;\;
\frac{k_0^W}{\Lambda'^2} = -0.000142\,TeV^{-2};\label{ak1}\\
& &\frac{a_0^W}{\Lambda'^2} = -1.044\cdot 10^{-32}\,TeV^{-2}\,;\;
\frac{k_0^W}
{\Lambda'^2} = -1.13\cdot 10^{-32}\,TeV^{-2}.\label{ak2}
\end{eqnarray}
Comparing the two last expressions and taking from
experimental work~\cite{19}
the following limitations
\begin{equation}
-21\, TeV^{-2}<\frac{a_0^W}{ \Lambda'^2}<20\,TeV^{-2};\;
-12\, TeV^{-2}<\frac{k_0^W}{ \Lambda'^2}<10\,TeV^{-2};
\label{eq:CMS2}
\end{equation}
we see, that predictions~(\ref{ak1}, \ref{ak2}) are
deeply inside boundaries of limitations~(\ref{eq:CMS2}).
The most recent limitations \cite{20} at $7\,-\,8\,TeV$, which essentially improve results for $a_0^W$
\begin{equation}
-1.1\, TeV^{-2}<\frac{a_0^W}{ \Lambda'^2}<1.1\,TeV^{-2};\;
\label{eq:CMSa2}
\end{equation}
also do not contradict estimates~(\ref{ak1}, \ref{ak2}).
Of course, the second solution~(\ref{ak2})
 gives a negligible small value, whereas the first one~(\ref{ak1})
 for a possibility of its checking
needs further essential improvement of the precision.

The second solution~(\ref{sol6}) of the set of compensation
equations gives the following solution
\begin{eqnarray}
& &x_2\,=\,x_4\,=\,0;\;x_3\,=\,-4.21777;\;x_5\,=\,-\,5.95333;\;
a\,=\,-\,0.87338;\label{solution51}\\
& &\Lambda\,=\,0.3646\,TeV;\;G_2 = G_4 = 0;\;
G_3 =  -\frac{1670}{TeV^2};\;G_5= -\frac{2360}{TeV^2}.\nonumber
\end{eqnarray}
The third solution~(\ref{sol9}) of the set of compensation
equations gives two solutions with
the same cut-off. We have the following sets of parameters
\begin{eqnarray}
& &x_2\,=\,x_4\,=\,-\,1.72596;\;x_3\,=\,x_5\,=\,3.9589;\;
a\,=\,-\,0.876955;\label{solution31}\\
& &\Lambda\,=\,0.1068\,TeV;\;G_2 = G_4 = -\frac{7970}{TeV^2};\;
G_3 = G_5= \frac{18270}{TeV^2};\nonumber\\
& &x_2\,=\,x_4\,=\,-\,0.864885;\;x_3\,=\,x_5\,=\,-2.61273;\;
a\,=\,0.876955;\label{solution32}\\
& &\Lambda\,=\,0.1068\,TeV;\;G_2 = G_4 = -\frac{3992}{TeV^2};\;
G_3 = G_5=-\frac{12060}{TeV^2};\nonumber
\end{eqnarray}
Solutions~(\ref{solution51}, \ref{solution31}, \ref{solution32})
evidently contradict limitations~(\ref{eq:CMS2}, \ref{eq:CMSa2}) due to very low value
for
cut-off $\Lambda$.

There is also solution~(\ref{solutionhigh}) with very large cut-off
$\Lambda$.
It is remarkable, that this solution
correspond to the cut-off being of the order of magnitude of the
Planck mass $M_{Pl} = 1.22\times 10^{16}\,TeV$. Of course effective
coupling constants $G_i$ in this case are extremely small.
This possibility in case of its realization may serve as an explanation
of hierarchy problem \cite{21}. Indeed, with this solution the actual values for the masses of
$W,\,Z,\,H$ and value $\alpha(M_Z)$ may be reconciled with the effective
cut-off being defined by the gravitational Planck mass. So the actual relation
between the electro-weak
scale and the gravity scale may acquire at least a qualitative interpretation.

We would draw attention to the low cut-off case also. Value of
$\Lambda$~(\ref{solutionlow}) is close to boundary value of the
momentum in the problem of a spontaneous generation of
anomalous triple $W$ interaction~(\ref{FFF}). Indeed, value of
the electro-weak gauge constant $g$ at this boundary~(\ref{eq:gz0})
\begin{equation}
g(\Lambda)\,=\,0.60366\,.\label{glambda}
\end{equation}
Then the following relation is to be fulfilled
\begin{equation}
\frac{g(\Lambda)^2}{4\,\pi}\,=\,\alpha_{ew}\,;\label{alphaew}
\end{equation}
where $\alpha_{ew}$ is defined in~(\ref{eq:a}). Then this relation
is an equation for  parameter $\Lambda$. The solution of this
equation gives
\begin{equation}
\Lambda\,=\,7.91413\cdot 10^{2}\,TeV\,.\label{Lambda}
\end{equation}
We see, that this value is of the same order of magnitude as
value $5.2262\cdot10^2\,TeV$ in solution~(\ref{solutionlow}).

Now we could formulate results in a rather different manner.
We have two interesting values for possible cut-off $\Lambda$.
The low value~(\ref{Lambda}), which is compatible with previous
results~\cite{9,10} by the order of magnitude, and the Planck mass.
Let us consider set of equations~(\ref{sol1}, \ref{BSW2}, \ref{BSW4})
for these
values of the cut-off. Earlier we have fixed actual value for
electromagnetic constant $\alpha(M_Z)$ and calculated values for
the cutoff~(\ref{solutionlow}, \ref{solutionhigh}). Now we fix
$\Lambda$ and calculate $\alpha(M_Z)$. In this way  for values
(\ref{Lambda}) and the Planck mass we obtain respectively
\begin{equation}
\alpha(M_Z)_{44}\,=\,0.00792;\quad
\alpha(M_Z)_{Pl}\,=\,0.00790.\label{alpha2}
\end{equation}
Both values differ from actual value $\alpha(M_Z)\,=\,0.007756$
by 2\%. Thus it might be possible to interpret results~(\ref{alpha2})
as a calculation of a value of
$\alpha$ with this precision. Just contributions of order of
magnitude of few \% are expected at the next approximation in the
development in powers of $\alpha_{ew}$.

Of course, there is the
trivial solution of set~(\ref{eq:CompWB}): all $x_i\,=\,0$, which
gives no additional information.  However we have also quite
informative non-trivial solutions.

The problem of the choice of the genuine solution is of course
essential. The answer is to be connected with the problem of a
stability of solutions. There are also possibilities of phase transitions
between different solutions. These problems are very difficult and need
extensive additional studies.

\section{Experimental implications}

Effective interaction~(\ref{eq:intG}) leads to effects in inclusive
reactions
\begin{equation}
p + p \to W^+ + W^- + W^\pm (Z, \gamma) + X. \label{pp3W}
\end{equation}
Unfortunately  with values for effective coupling constants
$G_2,\,G_4$ for preferable
solution~(\ref{solutionlow}, \ref{solutionlowG}) one
could hardly hope for achieving the necessary precision even
at the upgraded LHC.

However there is a possibility for an enhancement of the effect in
processes involving $t$-quarks due to large $m_t$. Let us consider the
wouldbe contribution of
interaction~(\ref{eq:intG}, \ref{solutionlow}) to vertex
\begin{equation}
\frac{G_{W \bar t t}}{2}\,\bar t  t \,W^b_{\mu \nu}\,W^b_{\mu \nu}\,.\label{ttWW}
\end{equation}
\begin{figure}
\centering
\includegraphics[scale=0.50]{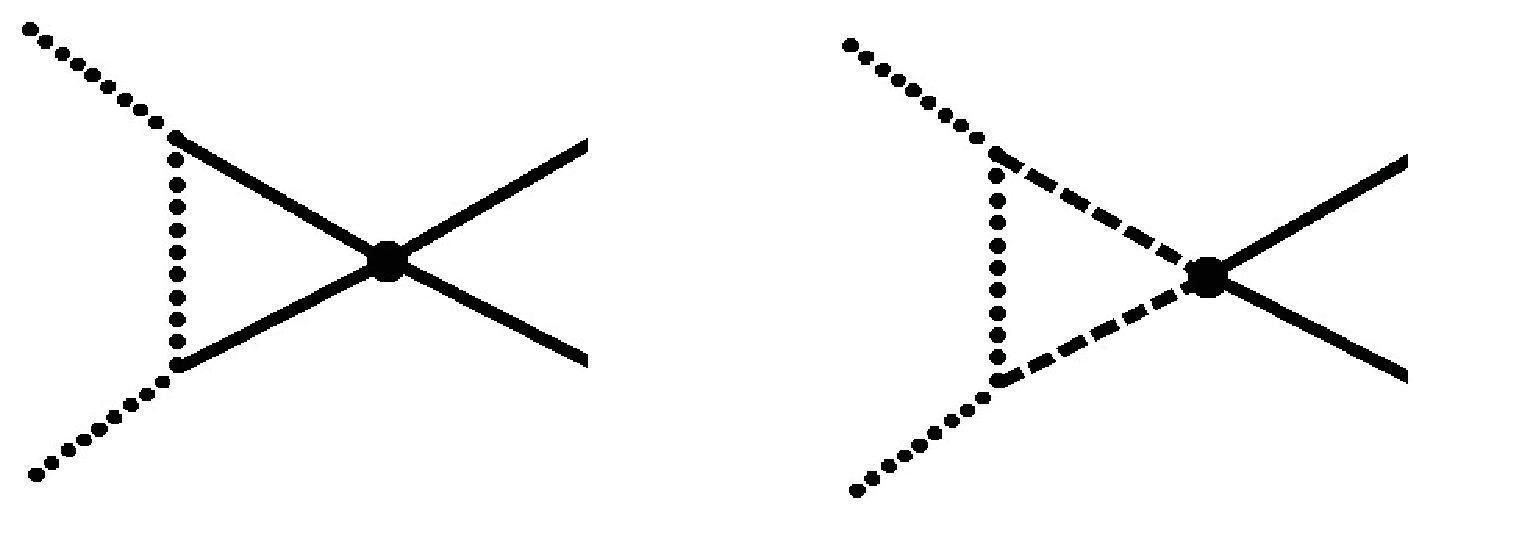}
\caption{Diagram representation of $\bar t t W W$ vertex. Dotted lines represent Z-bosons, simple lines represent $W^b$. The $t$-quarks are on the left.}
\label{fig:WWTt}
\end{figure}
The effective coupling for this
vertex is defined by diagrams presented in Fig. \ref{fig:WWTt}. We are
to
use the same cut-off $\Lambda$~(\ref{solutionlow}) in calculation
of this diagram and we obtain
\begin{equation}
G_{W \bar t t}\,=\,-\frac{g^2(\Lambda)\,M_t(\Lambda)}{24\,
M_W^4} \Bigl(2\,x_2+a^2(\Lambda)\,x_4\Bigr)\,=\,4.25 \cdot 10^{-8}
\,GeV^{-3};
\label{ttW}
\end{equation}
where we take parameters~(\ref{solutionlow}, \ref{glambda}) and for
$M_t(\Lambda)$ we use standard evolution expression
\begin{equation}
M_t(\Lambda)\,=\,\frac{M_t}{\biggl( 1 + \frac{7 \alpha_s(M_t)}
{4 \pi} \ln \biggl[\frac{\Lambda^2}{M_t^2}\biggr]\biggr)
^{\frac{4}{7}}}
;\label{mtlam}
\end{equation}
where $M_t = 173.2\,GeV$ is the table value for the
$t$-quark mass \cite{15}.
Let us consider processes $p+p\to\bar t\,t\,W^+(Z)$.
With value~(\ref{ttW}) we have an additional contribution of
the new effective interaction~(\ref{ttWW}) to the cross section
$\sigma_{\bar t t W}$ of process\footnote{We have got persuaded,
that an interference of contributions of effective
interaction~(\ref{ttWW}) with the SM terms is negligible.}.
\begin{equation}
p + p\,\to\,\bar t + t +(W^\pm,Z)+X;\label{reactW}
\end{equation}
for $\sqrt{s} = 8\,TeV$
the following estimate
\begin{equation}
\Delta \sigma_{\bar t t W^+}(8\,TeV)\,=\,103.5\, fb.\label{8}
\end{equation}
For the same process with the negative $W$ we have
\begin{equation}
\Delta \sigma_{\bar t t W^-}(8\,TeV)\,=\,28.0\, fb.\label{8-}
\end{equation}
For process $p + p\,\to\,\bar t + t +Z $ we have the following
contribution
\begin{equation}
\Delta \sigma_{\bar t t Z}(8\,TeV)\,=\,47.2\, fb.\label{8Z}
\end{equation}
These results, as well as the subsequent ones, are obtained with
use of the CompHEP package \cite{22}.

Recent CMS result at $\sqrt{s} = 8\,TeV$ \cite{23} for these processes\footnote{Results
for $\sqrt{s}=7\,TeV$ see in \cite{24}.}
\begin{eqnarray}
& &\sigma_{\bar t t W^+}(8\,TeV)\,=\,170^{+110}_{-100}\, fb;
\label{CMStt}\\
& &\sigma_{\bar t t Z}(8\,TeV)\,=\,200\pm 90\, fb;
\nonumber
\end{eqnarray}
Results~(\ref{CMStt}) are
compatible with wouldbe additional
contributions~(\ref{8}, \ref{8Z}) as well as with the Standard Model.
There is no data for process~(\ref{8-}) in~\cite{23}.

\bigskip

\begin{table}[!h]
\caption{ SM results for cross-sections of processes $p+p\to\bar t t V$ at
$\sqrt{s}=8 TeV$ and $\sqrt{s}=14 TeV$ and predictions for
additional contributions due to effective interaction~(\ref{ttWW}).}
\label{table1}
\centering
\begin{tabular}{||c|c|c|c|c||}
\hline
channel & $\sigma_{SM}\,fb,\,8\,TeV $ & $\Delta\sigma\,fb,\,
8\,TeV$
&
$\sigma_{SM}\,fb,\,14\,TeV $ & $\Delta\sigma\,fb,\,14\,TeV$ \\
\hline
$\bar t t W^+$ & $161^{+19}_{-32}$  & 103.5 & $507^{+147}_
{-111}$ & 1257\\
\hline
$\bar t t W^-$ & $ 71^{+11}_{-15}$ & 28.0  &
$ 262^{+81}_{-60}$
& 355\\
\hline
$\bar t t Z$ & $197^{+22}_{-25}$ & 47.2  &
$760^{+74}_{-84}$& 578\\
\hline
\hline
\end{tabular}
\end{table}

There are the most recent data at $\sqrt{s} = 8\,TeV$ of both ATLAS~\cite{28}
and CMS~\cite{29} collaborations:
\begin{equation}
t\,\bar t\,W: \sigma = 369^{+100}_{-91}\,fb;\quad
t\,\bar t\,Z: \sigma = 176^{+58}_{-52}\,fb;\;(ATLAS)\label{ATLASNEW}
\end{equation}
\begin{equation}
t\,\bar t\,W: \sigma = 382^{+117}_{-102}\,fb;\quad t\,\bar t\,Z: \sigma = 242^{+65}_{-56}\,fb;\;(CMS)\label{CMSNEW}
\end{equation}
For process $pp\to\,t \bar t\,W$ with both charge signs we have in new
data results, which agree both the SM value $\simeq 232\,fb$
and the predicted one~$\simeq 363\,fb$ . However one sees, that data of both collaborations slightly favor the last predicted value.

Let us note, that additional contributions $\Delta \sigma(\bar t t W,Z)$ increase
with the energy and for the updated energy of the LHC
$\sqrt{s} = 14\,TeV$ they become
\begin{eqnarray}
& &\Delta \sigma_{\bar t t W^+}(14\,TeV)\,=\,1257\, fb.\label{14}\\
& &\Delta \sigma_{\bar t t W^-}(14\,TeV)\,=\,355\, fb.\label{14-}\\
& &\Delta \sigma_{\bar t t Z}(14\,TeV)\,=\,578\, fb.\label{14Z}
\end{eqnarray}
Our predictions are to be compared with the SM
calculations~\cite{25,26,27} in Table~\ref{table1}\footnote{The result
for $\sigma_{SM}(\bar t t Z)$ in the fourth column
corresponds to $\sqrt{s}=13\,TeV$.}.

We have already noted, that results for $\sqrt{s} = 8\,TeV$ do not
contradict the current data~(\ref{CMStt}, \ref{ATLASNEW}, \ref{CMSNEW}). As for
$\sqrt{s} = 14\,TeV$,
we see from the Table, that the most promising process for testing
of the present results
at the upgraded LHC is $p + p \to \bar t\,t\,W^\pm$. Indeed,
the total additional contribution to production of charged $W$
with top pair
is around $1.6\,pb$, that more than twice exceeds the corresponding
total SM value. Note, that we do not include in the Table
process $p + p \to \bar t\,t\,\gamma$, because the effect here is
significantly less pronounced. Namely, for $\sqrt{s} = 13\,TeV$
we have $\sigma_{SM} = 1.744 \pm 0.005\,pb$~\cite{27}, whereas
the effect of interaction~(\ref{ttWW}) is calculated to be
$\Delta \sigma = 0.125\,pb$. Let us also note, that our estimations
of the effect might have accuracy around $10\%$ according to the
experience of applications of the approach to several examples
(see book \cite{12}).

Provided the prediction being confirmed, the first
non-perturbative effect in the electroweak
interaction would be established.

\section{Conclusion}

To conclude let us draw attention to the the results in view of
the compensation approach to the problem of a spontaneous generation
of an effective interaction. First of all, the results are obtained
exclusively due to application of this approach. We would emphasize
that the existence of a non-trivial solution of compensation
conditions always impose strong restrictions on parameters of the
problem. We see such restrictions in both problems of the spontaneous
generation of the Nambu -- Jona-Lasinio interaction \cite{7} and
the triple anomalous weak boson interaction \cite{9,10}. In the present work such conditions for existence of
interaction (\ref{eq:LeffWZ}) are shown to define the Weinberg mixing angle,
that leads to results (\ref{alpha2}) for the electromagnetic coupling constant.

It is also worth mentioning, that the wouldbe effective interaction under consideration leads to significant experimental effects being shown in
Section 3, which likely may be either proved or disproved in forthcoming studies at the LHC.

\section{Acknowledgements}
The work is supported in part by grant NSh-7989.2016.2 of the President of Russian Federation.

\end{document}